\documentclass[aps,pra,twocolumn,superscriptaddress,floatfix]{revtex4}

\usepackage{bm}
\usepackage[colorlinks=true, citecolor=blue, urlcolor=blue]{hyperref}\usepackage{epsf,graphics,graphicx}
\usepackage{amsmath}
\usepackage{latexsym}
\usepackage{amssymb}
\usepackage{graphicx}
\usepackage{color}
\usepackage{soul,xcolor}
\setstcolor{red}
\usepackage{ulem}
\usepackage{cancel}
\usepackage{tikz}
\usepackage[export]{adjustbox}
\usepackage{subcaption}
\usepackage{braket}

\newcommand{\beq}{\begin{equation}}
\newcommand{\eeq}{\end{equation}}

\usepackage{epsfig}

\input{epsf}

\begin{document}
\title{Photo-induced tunable Anomalous Hall and Nernst effects in tilted Weyl Semimetals using Floquet theory}
\author{Anirudha Menon}
	\affiliation{Department of Physics, University of California, Davis, California 95616, USA} \email{amenon@ucdavis.edu}
	\author{Debashree Chowdhury}
	\affiliation{Physics Department, Ben Gurion University, Beer Sheva 84105, Israel} \email{debashreephys@gmail.com}
	\author{Banasri Basu}
	\affiliation{Physics and Applied Mathematics Unit, Indian Statistical Institute, Kolkata 700108, India}
	\email{sribbasu@gmail.com}

\date{\today}

\begin{abstract}
In this paper, we discuss the effect of a periodically driving circularly polarized laser beam in the high frequency limit, on the band structure and thermal transport properties of type-I and type-II Weyl semimetals (WSMs). We develop the notion of an effective Fermi surface stemming from the time-averaged Floquet Hamiltonian and discuss its effects on the steady-state occupation numbers of electrons and holes in the linearized model. In order to compute the transport coefficients averaged over a period of the incident laser source, we employ the Kubo formalism for Floquet states and show that the Kubo formula for the conductivity tensor retains its well known form with the difference that the eigenstates and energies are replaced by the Floquet states and their quasi-energies. We find that for type-I WSMs the anomalous thermal Hall conductivity grows quadratically with the amplitude $A_{0}$ of the U(1) gauge field for low tilt, while the Nernst conductivity remains unaffected. For type-II WSMs, the Hall conductivity decreases non-linearly with $A_{0}$ due to the contribution from the physical momentum cutoff, required to keep finite electron and hole pocket sizes, and the Nernst conductivity falls of logarithmically with $A_{0}^2$. These results may serve as a diagnostic for material characterization and transport parameter tunability in WSMs, which are currently the subject of a wide range of experiments.

\end{abstract}
\maketitle

Weyl Semimetals (WSMs) have been identified as materials with nontrivial topological structure having a variety of physical properties stemming from a Hamiltonian possessing a gapless spectrum with at least one of the time reversal and inversion symmetries broken \cite{1,2,14,18}. The minimal model, obtained by breaking time reversal symmetry, consists of a Dirac-like dispersion around two distinct points in the first Brillouin zone, where the conduction and the valence bands touch. These Weyl points or Weyl nodes are topological charges acting  as a source or a sink for Berry curvature \cite{2,16,19}, as reflected by their occurrence in opposite chirality pairs, and by contrast, the inversion symmetry breaking minimal model requires four Weyl points \cite{1,18}.  Such materials, classed as type-I WSMs, exhibit a number of phenomena including chiral magnetic waves \cite{Khar}, chiral anomaly induced plasmon modes \cite{JZ}, and chirality induced negative magneto resistance \cite{Nin}. 
The addition of a SO(3,1) symmetry breaking term to the low energy Hamiltonian coupled to the momentum leads to a tilt in the dispersion. For sufficiently large tilts, it can be shown \cite{1,8} that a Lifshitz phase transition occurs, leading to a new phase i.e. type-II WSMs, with different physical properties. Type-I WSMs have a single Fermi surface, whereas in type-II WSMs, the Fermi surface splits into two, one each for electrons and holes, such that the density of states at each Weyl points is finite. Reports on the experimental realizations of type-I Weyl semimetals have been presented in \cite{hasan,a}, and it was shown in \cite{Wte} that $\textrm{WTe}_2$  is a possible experimental candidate for the type-II WSM phase. 

Light-matter interactions allow for an effective mechanism to create steady state exotic phases of matter, and to this end, Floquet engineering is a novel approach which can be implemented on ultrafast timescales. The essence of Floquet theory lies in simplification of a time dependent problem to an effective time independent form, and it is used to diagnose the behavior of physical observables in different phases \cite{Tahir,debabrata,Tahir1}. The use of a high frequency ($\omega$) laser field, which is not directly involved in any electron transitions, but is associated with the virtual photon absorption and emissions, leads to a plethora of interesting effects in photonic crystals \cite{22}, graphene \cite{29}, silicene \cite{Ezawa} and topological insulators \cite{23,24,Zhongbo,Torres,Torres 1,Torres 2,Torres 3,Cayssol,Ours}. Additionally, the high frequency limit (HFL) permits a perturbative expansion of the infinite dimensional Floquet Hamiltonian in powers of $1/\omega$ [the high frequency expansion (HFE) or Van-Vleck expansion], leading to a finite dimensional effective Hamiltonian which helps preserve computational tractability. 

These reports motivate the study of irradiated WSMs in the HFL \cite{10,3} and the effects generated on the transport coefficients with a goal of obtaining tunable handles on properties such as Hall conductivity, anamolous thermal Hall conductivity and Nernst conductivity. The thermal transport coefficients of WSMs carry signatures of exotic physics like the chiral anomaly and Berry curvature, generating a significant amount of recent experimental interest to characterize such materials \cite{Hir,Lv,Gooth}, to find potential applications in nanodevices. Subtleties regarding the applicability of the linearized minimal model close to the Lifshitz transition \cite{1} have been discussed here, and we have taken care to conduct our analyses in the heart of either WSM region. A stable WSM phase can obtained by irradiating a Dirac semimetal with circularly polarized light \cite{5}, and we show that this allows for controllable Weyl node separation and effective Fermi surfaces, leading to tunable off-diagonal transport coefficients.\\

Consider a time reversal symmetry breaking tilted Weyl semimetal with two Weyl nodes of opposite chirality. The  linearized Hamiltonian for such a system around each Weyl node $s=\pm$ is given by \cite{15}
\begin{align} 
H_{s}=\hbar C_{s}(k_z-sQ)+s\hbar v\bm\sigma\cdot(\bm k-sQ\bm e_z)
\label{1}
\end{align}
where $C_{s}$ is the tilt parameter, which also is associated with the type of the Weyl point. Here, $v$ denotes the Fermi velocity in the absence of the tilting term, $2Q$ is the distance between the Weyl points in momentum space along $\bm e_z$, and $\bm \sigma$ is the vectorized Pauli matrix. 

\begin{widetext}

\begin{figure} [h]
\centering
\begin{subfigure}[t]{.4\textwidth}
\centering
\fbox{\includegraphics[scale = .9]{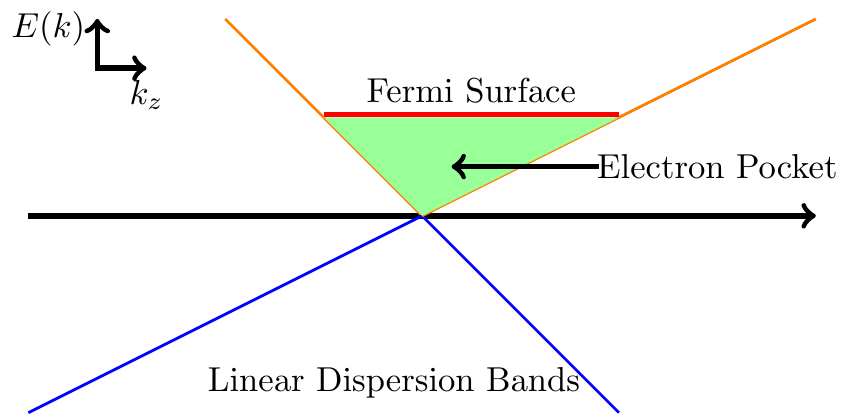}}
\caption{} 
\end{subfigure}
\hfill
\begin{subfigure}[t]{.5\textwidth}
\centering
\fbox{\includegraphics[scale=.9]{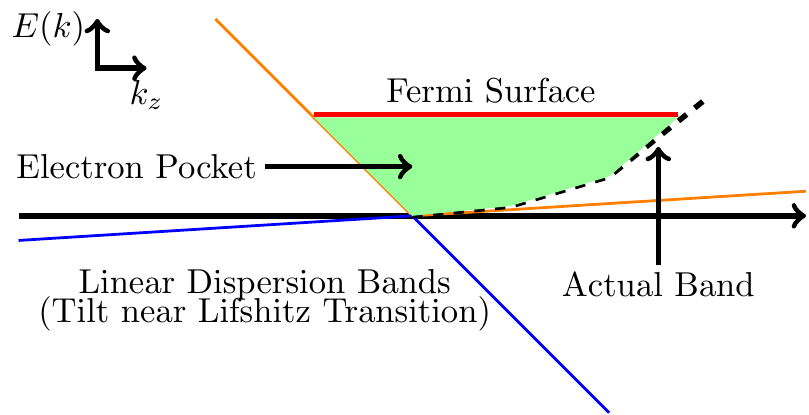}}
\caption{} 
\end{subfigure}

\begin{subfigure}[t]{.35\textwidth}
\centering
\fbox{\includegraphics[scale = .81]{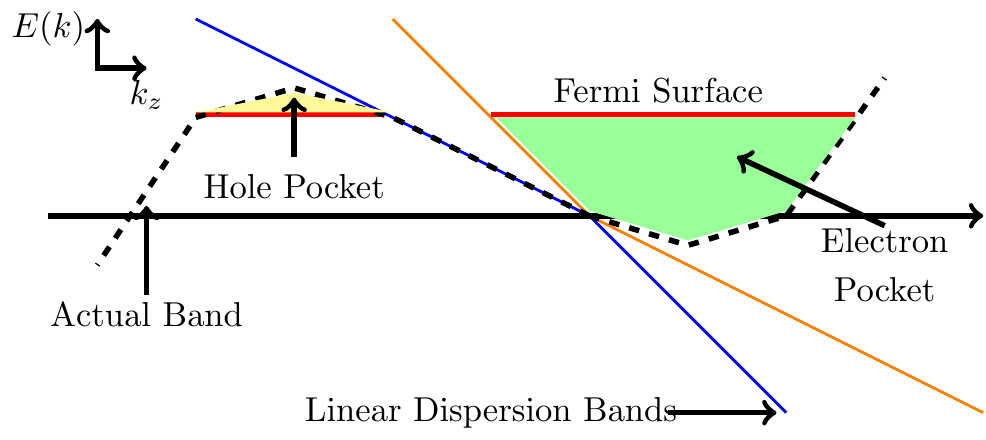}}
\caption{} 
\end{subfigure}
\hfill
\begin{minipage}[b]{.45\textwidth}
\captionof*{figure}{FIG. 1: (a) Fermi surface relative to Weyl node for type-I WSM away from the Lifshitz transition. The upper band is in orange and the lower band is in blue. (b) Type-I WSM - failure of the linearized model near the Lifshitz transition. (c) Fermi surface relative to Weyl node for type-II WSM showing how the linearized model can give us qualitatively correct results by imposing a physical momentum cutoff since the actual band structure (dashed lines) have finite electron and hole pockets.}
\end{minipage}
\end{figure}

\end{widetext}

For type-I WSMs shown in Figs. 1(a) and 1(b), the blue and orange lines indicate the linearized band structure near the Weyl nodes (with their meeting point being the Weyl point), the red lines indicate the Fermi energy or highest occupied level, and the green pockets indicate the zone filled by electrons. It is clear from Fig. 1(b) that as we increase the tilt the higher order momentum terms in the Hamiltonian become relevant and while the actual electron pocket size is finite as indicated by the the green zone (the dashed boundary corresponds to the actual band structure with higher order corrections), the linearized model predicts infinite electron pocket sizes. Fig. 1(c) shows that for type-II WSMs, past the Lifshitz transition, a physical momentum cutoff needs to be introduced since the true band structure admits only finite pocket sizes. \\

We use a polarized beam of form $E(t)=E_{0}(\cos \omega t,-\sin \omega t)$, where $E_{0}$ and $\omega$ are the amplitude and frequency of the optical field. The Pierels substitution leads to $\hbar k_{i} \rightarrow \hbar k_{i}+eA_{i}$, where $\vec{A}(t+T)=\vec{A}(t)$, with $T=2\pi/ \omega$ as the periodicity. The full time-dependent Hamiltonian has the form

\begin{align}\label{555} 
H_s(k,t) &= H_0(k) +\mathcal{V}_s(t)
\end{align}
with $H_0(k)=H_s= \hbar C_{s}(k_z-sQ)+s\hbar v\bm\sigma\cdot(\bm k-sQ\bm e_z)$ and $\mathcal{V}_s(t)= s\hbar \nu A_0(\sigma_x \sin \omega t +\sigma_y \cos\omega t)$.
In the HFL, we map to a time-independent problem by using Floquet theory and employ the HFE \cite{debabrata,100,200,41,awe}, for which a brief review is provided in the supplementary material (Appendix A).  \\ 

The effective time independent Hamiltonian for our system can be obtained as,
\begin{align}\label{7}
H_{F}^{s} &=\hbar C_{s}[k_{z}-s(Q+\Delta)]+s\hbar v\bm\sigma\cdot[\bm k-s(Q+\Delta)\bm e_z]\nonumber\\&+s \hbar C_{s} \Delta,
\end{align}
where $\Delta =\frac{\hbar  v A_{0}^{2}}{2\omega},$ is the contribution of the radiation field. It is to be noted here from eqn.(\ref{7}) that the form of the effective Hamiltonian is similar to the original Hamiltonian in eqn.(\ref{1}), with the Weyl nodes being further displaced by a distance $2\Delta$ in momentum space.  We restrict to the inversion symmetric case, $sC_s = C, \forall s = \pm$, and so there is an overall shift in the total energy of both nodes by an amount equal to $\hbar C \Delta$. \\ \\

In order to analyze the Nernst conductivities \cite{6} and anomalous thermal Hall conductivities \cite{7,9,12} in both the regimes of WSMs, we can now use the Kubo formalism modified for Floquet theory \cite{PHG,Torres,RefC}. The modification to the standard form of the Kubo formula, used to calculate the time-averaged conductivity tensor for periodically driven systems, lies in the use of Floquet states, quasi-energies, and the time averaging integral, as expained in Appendix B. We show in Appendix B (eqn. B10) that for a Hamiltonian linear in momentum, this can be simplified to the form
\begin{widetext}
\beq
\sigma_{ab} = i \int \frac{d{\bf k}}{(2\pi)^3} \sum_{\alpha \neq \beta} \frac{f_\beta({\bf k}) - f_\alpha ({\bf k})}{\epsilon_\beta ({\bf k}) - \epsilon_\alpha ({\bf k})} \times \frac{ \braket{e_\alpha  ({\bf k}) | J_b | e_\beta  ({\bf k}) } \braket{e_\beta  ({\bf k}) | J_a | e_\alpha  ({\bf k}) }}{\epsilon_\beta ({\bf k}) - \epsilon_\alpha ({\bf k}) + i\eta },
\eeq  
\end{widetext}
which resembles exactly the standard form of the Kubo formula where $J_a(b)$ represents the current operator, the $e_\alpha$'s represent the states of the effective Floquet Hamiltonian, and the $\epsilon$'s represent the corresponding quasi-energies. The $f_{\alpha}$'s represent the occupations which in general could be non-universal in systems which are out of equilibrium. However, the steady-state occupations resemble the Fermi-Dirac distribution with the quasi-energies for a certain class of system-bath couplings \cite{Eqb}, and our results are valid for such cases.\\ \\

In general, Mott's relationship defines the anomalous thermal Hall and Nernst conductivities as \cite{6,7,9,12},
 \begin{gather}
\alpha_{xy}=eLT\frac{d\sigma_{xy}}{d\mu},\label{26}~~~
K_{xy}=LT\sigma_{xy},
\end{gather}
where  $L=\pi^2k_B^2/3e^2$ is the Lorentz number, $e$ is the electronic charge, and $k_B$ is the Boltzmann constanst. The calculation of $\sigma_{xy}$ using the Matsubara approach is detailed in the supplementary material (Appendix C). It is to be noted here that the electric field, sourced by the small d.c. bias needed to measure transport properties in the linear response regime, has been suppressed in eqn.(\ref{555}), and we account for it in the conductivity tensor calculation presented in Appendix C [eqn.(C2)].
From eqn.(\ref{26}), the Nernst and thermal Hall conductivities in the $T \rightarrow 0$ limit for a type-I WSM, are obtained as,

\begin{align}\label{27}
\alpha_{xy}&=\frac{-ek_{B}^{2}TC}{18\hbar^{2}v^{2}}\nonumber\\
K_{xy}&\approx \frac{k_{B}^{2}T}{6\hbar}\Big[\Big(Q+\Delta\Big)-\frac{C(\mu-C\Delta)}{3\hbar v^{2}}\Big].
\end{align}

\begin{figure} [h]
\centering
\fbox{\includegraphics[scale = .9]{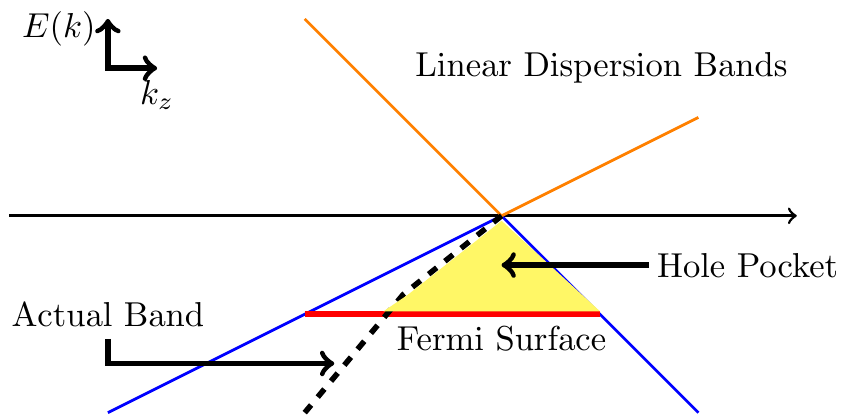}}
\caption{Fermi surface relative to Weyl node for type-I WSM showing that on lowering the effective chemical potential we eventually get incorrect estimates of the hole pocket size.}
\end{figure}

Firstly, we note that $K_{xy}$ varies smoothly around the point $\mu = C\Delta$, and that setting $\Delta = 0$ gives us back the results in \cite{1}. We rewrite the Hall conductivity in a suggestive manner which shows that the Hall conductivity grows monotonously with $\Delta$. 

\beq K_{xy} = \frac{k_B^2 T}{6 \hbar} [ Q - \frac{\mu }{3 \hbar v^2} C] + \frac{k_B^2 T}{6 \hbar} [1 +  \frac{C^2}{3 \hbar v^2} ] \Delta = K_{xy}^0 + K_{xy}^{\Delta} ,\eeq

where $K_{xy}^0$ is the Hall conductivity in the absence of irradiation and $K_{xy}^{\Delta}$ is the positive contribution of the laser field. Since both nodes get an energy boost of $C \hbar \Delta$, the chemical potential which shows up in the Fermi-Dirac distribution function is offset by it, and it's instructive to think of this as fixing the chemical potential and moving the band structure for both nodes vertically. It's clear that as we increase the amplitude of the radiation field, the effective chemical potential $\mu-C\Delta$ decreases and ultimately becomes negative. Since moving the effective chemical potential further down amounts to increasingly incorrect hole pocket size estimations in the linearized model (Fig. 2), with the dashed line indicating the actual band structure, one might worry about the limit of validity of the result. However, the free carrier contribution is a second order effect, supressed by $\frac{C^2}{v^2}$, and the dominant contribution to Hall conductivity comes from the shift in node spacing, i.e. $\Delta$, which is part of the vacuum contribution, known to be cutoff independent. Thus we can conclude that the Hall conductivity grows with the amplitude of the irradiation field far away from the linear regime. In Figs. 3(a) \& (b), the anomalous thermal Hall conductivity is plotted in units of $(k_{B}^{2}/\hbar)$ as a function of optical frequency and temperature.

Since the linearized model predicts a linear dependence of $K_{xy}$ on $\mu$ in the type-I regime, the Nerst conductivity is predictably constant and remains unchanged by the optical field.

\begin{widetext}

\begin{figure} [h]
\centering
\begin{subfigure}[t]{.4\textwidth}
\fbox{\includegraphics[scale = 0.54]{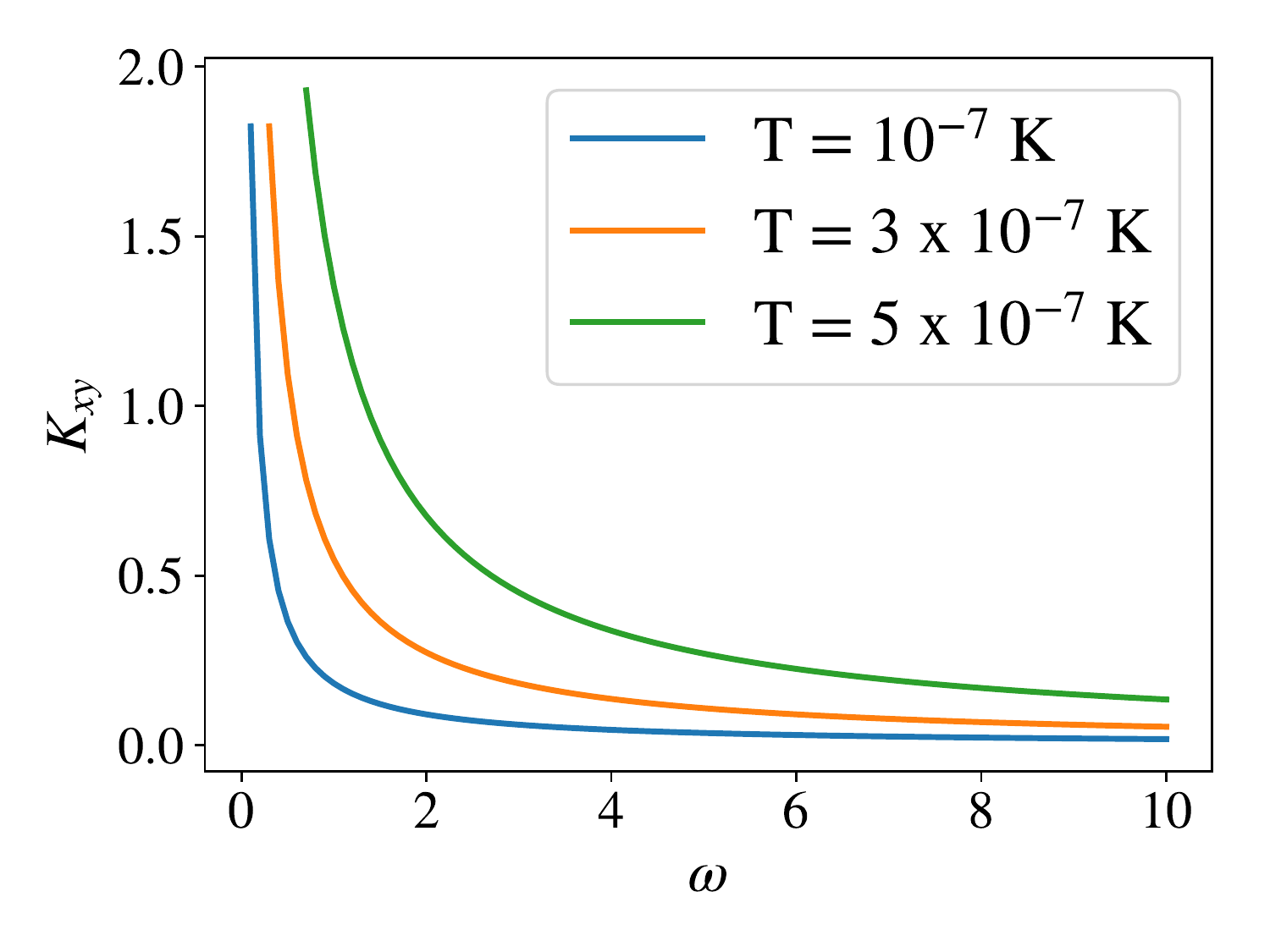}}
\caption{}
\end{subfigure}
\hfill
\begin{subfigure}[t]{.45\textwidth}
\fbox{\includegraphics[scale=0.54]{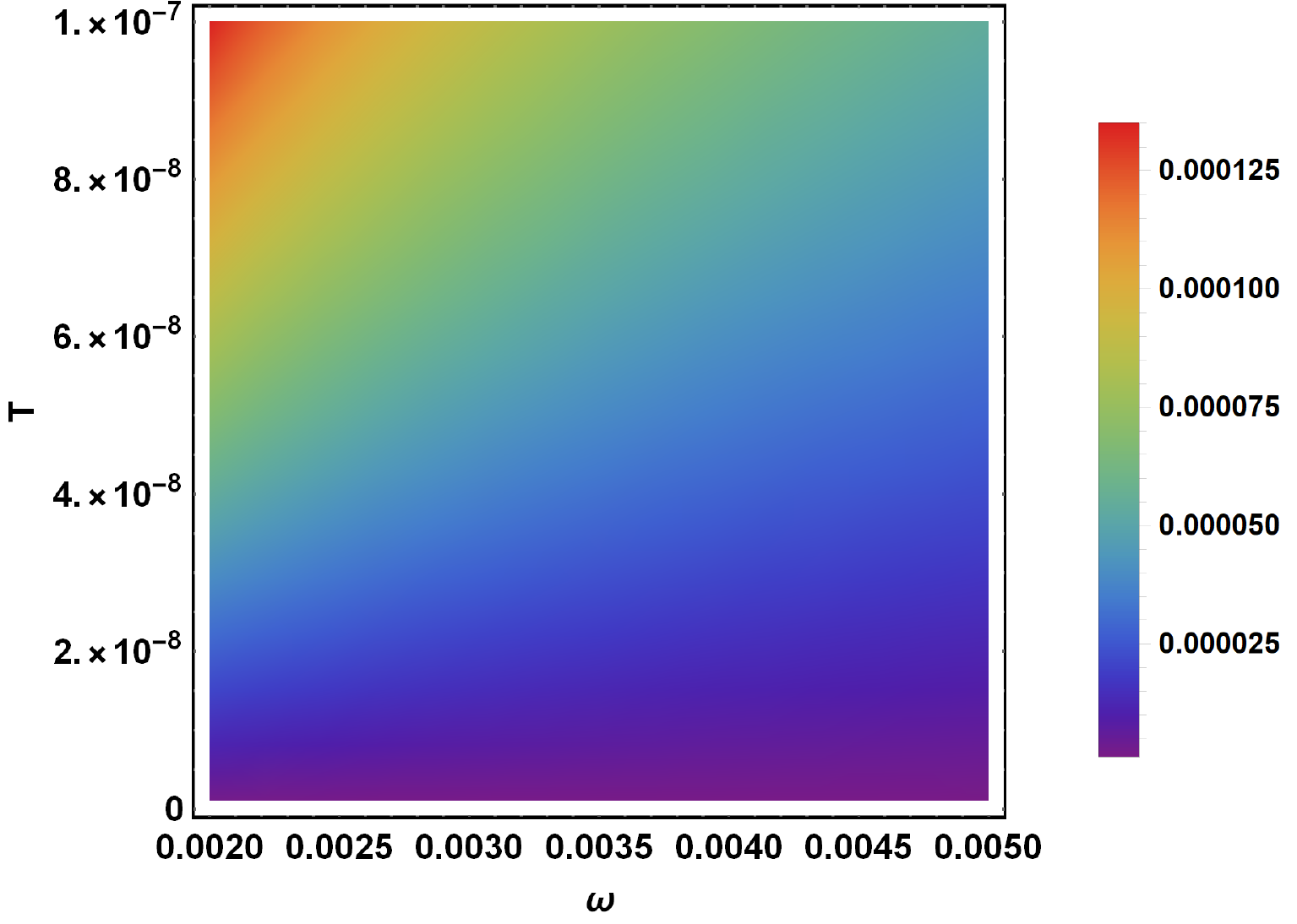}}
\caption{}
\end{subfigure}
\caption{(a) Variation of thermal Hall conductivity with optical frequency for three different values of the temperature for type-I WSM. The frequency axis is in units of $10^{12}$ Hz. (b) Variation of thermal Hall conductivity with optical frequency and temperature for type-I WSM. The frequency axis is in units of $10^{16}$ Hz, and the temperature axis is in units of $10^4 K$.}
\end{figure}
\end{widetext}

The Hall and Nernst conductivities for type-II WSMs are presented below.

\begin{align}\label{30}
\alpha_{xy}&=-\frac{ek_{B}^{2}T v}{6\hbar^{2}C^{2}}\Big{[}\ln\Big(\frac{C^{2}\Lambda}{v(C\Delta -\mu)}\Big)-1\Big{]}\nonumber\\
K_{xy}&=\frac{ek_{B}^{2}T v}{6C\hbar}\Big[\Big(Q+\Delta\Big)-\frac{(\mu-C\Delta)}{\hbar C}\ln \Big(\frac{C^{2}\Lambda}{v(C\Delta -\mu)}\Big)\Big].
\end{align}

Considering the expression for $K_{xy}$, which depends nonlinearly on the chemical potential,  it is clear that it decreases for increasing $\Delta$, and we find that changing the amplitude of the photon field affects the Nernst conductivity, which decreases logarithmically with increasing $\Delta$. For the correct qualitative description of the transport coefficients, the momentum cutoff needs to be modified for increasing $A_0$ .

Finally, while the physical momentum cutoff is difficult to estimate without using a non-linear model, we can provide a way to experimentally verify our findings independent of the cutoff. Notice that we can eliminate the $\Lambda$ dependence from eqns.(\ref{30}) to get:

\begin{align} \label{32}
[-\frac{6\hbar C}{k_B^2 T v} K_{xy} + Q +\Delta ] \frac{\hbar C}{C\Delta -\mu} &= \frac{6\hbar^2 C^2}{e k_B^2 T v} \alpha_{xy} +1
\end{align}

Figs. 4(a) \& 4(b) show the anomalous Hall conductivities for a range of Nernst conductivities and driving frequency at fixed temperature, as defined by eqn.(\ref{32}). \\ \\

A pump-probe experiment (Appendix D) is a potential candidate setup for the verification of the results stated here. Such setups have been used to create stable WSMs from Dirac metals and allow for the steering of Weyl points \cite{5}. Since the timescale for amplitude modulation is orders of magnitude larger than the oscillation of the field, the position of the Weyl nodes is dictated by the frequency of the optical field, with small variations due to amplitude modulation. Similar experiments have been also proposed for the transport properties of other driven topological phases \cite{Buc,Bang,Ed,23}. The effective Floquet band close to the Weyl node can be experimentally confirmed using time-resolved photo emission spectrosocopy \cite{70,5}. \\ \\

\begin{widetext}

\begin{figure} [h]
\centering
\begin{subfigure}[t]{.4\textwidth}
\centering
\fbox{\includegraphics[scale = 0.55]{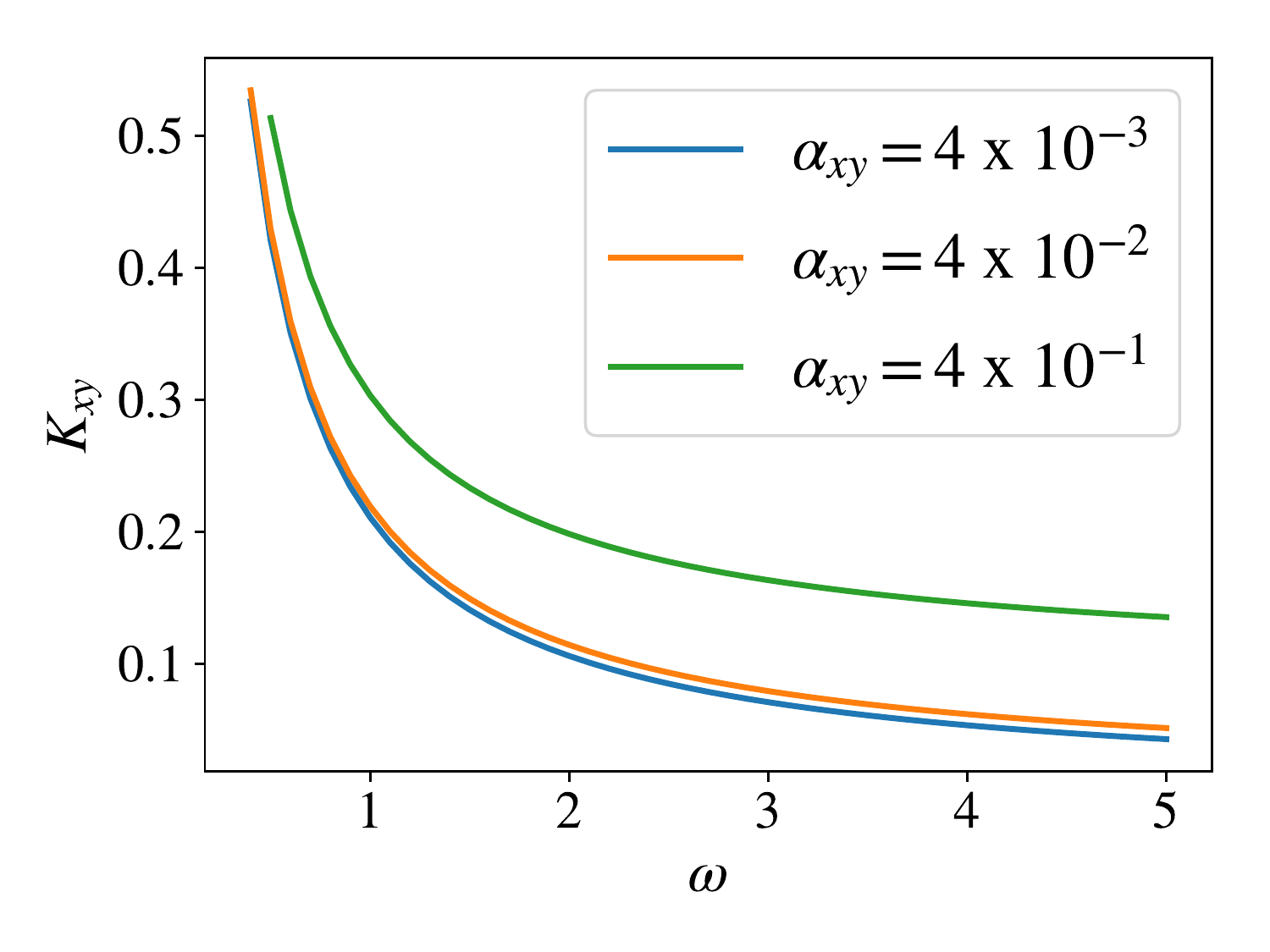}}
\caption{}
\end{subfigure}
\hfill
\begin{subfigure}[t]{.45\textwidth}
\centering
\fbox{\includegraphics[scale=0.65]{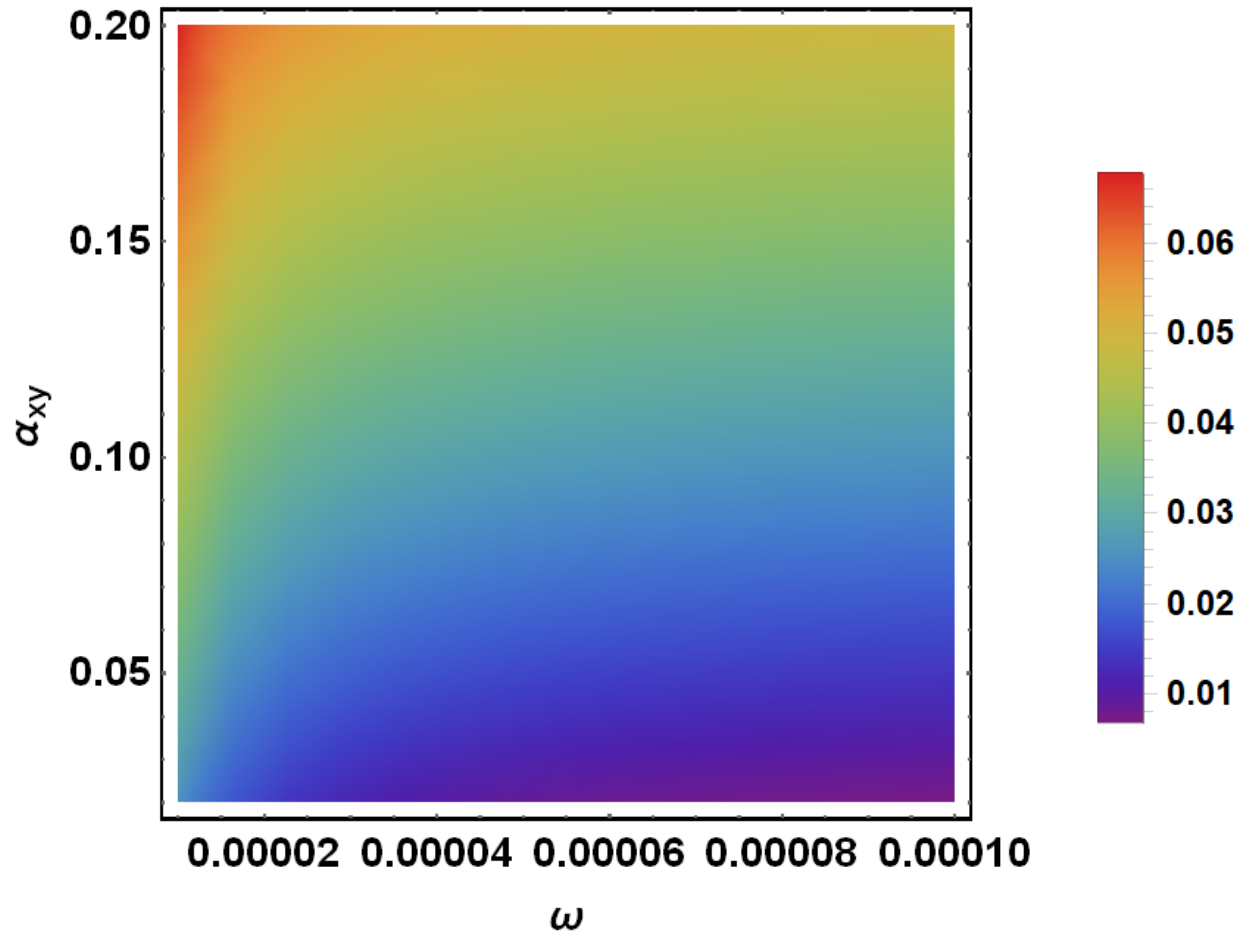}}
\caption{}
\end{subfigure}
\caption{(a) Variation of thermal Hall conductivity with optical frequency for three different values of Nernst conductivity for type-II WSM. The frequency axis is in units of $10^{10}$ Hz. (b) Variation of thermal Hall conductivity with optical frequency and Nernst conductivity for type-II WSM. The frequency axis is in units of $10^{16}$ Hz.}
\end{figure}

\end{widetext}

To conclude, in this letter we have considered the effects of an incident circularly polarized optical field on two distinct classes of Weyl Semimetals in the HFL using Floquet theory. The corresponding changes in thermal Hall conductivity and Nernst conductivity have been calculated for the linearized model, with closed form expressions for the $T \rightarrow 0$ case. These results and the underlying physics can be summed up as follows. 

For the effective Floquet Hamiltonian, we find that the Weyl nodes separate further due to the radiation field dependent parameter $\Delta$. This also gives rise to a constant term in the Hamiltonian proportional to $\Delta$, which leads to distinct shifts in the spectrum of each Weyl node and is shown to be equal in the inversion symmetric case. Thus, the effect of the latter is to change the effective Fermi surface leading to an array of consequences for the transport coefficients. The time-averaged transport coefficients are computed using the modified Kubo formalism applicable to Floquet states, and we show that the conductivity tensor can be computed using the Matsubara Green's function formalism, the key point being that the Kubo formula can be used in it's standard form with the states and energies in the expression being interpreted as the states and quasi energies of the effective Floquet Hamiltonian. 

For the type-I WSM case, we find that the leading correction to the Hall conductivity arises from the Floquet parameter $\Delta$. There exist subleading order corrections stemming from the true band structure which may not be accurately captured by the linearized model. The Nernst conductivity remains unchanged by the optical field because the Hall conductivty in the type-I regime shows a linear dependence on the chemical potential. 

In the type-II case, we find that the Hall conductivity decreases with the amplitude of the incident laser beam, holding the frequency fixed. The Nernst conductivity for this type of WSM is affected by the radiation field as the thermal Hall conductivity depends non-linearly on the chemical potential. With increasing $\Delta$, the Nernst conductivity falls off logarithmically. The qualitative and quantitative analyses of the transport properties of WSMs presented here aims to the characterization of the two types of WSMs. 

\begin{acknowledgements}
Work at BGU of D.C is supported by grants from Israel Science Foundation (ISF), the infrastructure program of Israel Ministry of Science and Technology under contract 3-11173, and the  Pazy Foundation. DC also acknowledges the financial support from the  Planning and Budgeting Committee (PBC).

The authors would like to thank the anonymous referees for their constructive comments and suggestions which have lead to the current form of the paper.
\end{acknowledgements}

\end{document}

\title{Supplementary Material: Photo-induced tunable Anomalous Hall and Nernst effects in tilted Weyl Semimetals using Floquet theory}
\maketitle

\appendix
\section{Floquet effective Hamiltonian}
For the sake of completeness we provide a brief review of the Floquet theory. The off resonant condition (i.e. the presence of only virtual photon processes)\cite{Tahir,Tahir1} is maintained here, which effectively makes the system stationary, and the static effective Hamiltonian in terms of the evolution operator $U$ \cite{41} is obtained as
\begin{eqnarray}
\mathcal{H}_{eff}(k)=\frac{i\hbar}{T}\log U ,
\end{eqnarray}
where,
\begin{align}
U =T_{time} \exp[\frac{1}{i\hbar}\int^{T}_{0}\mathcal{H}(k,t)dt]
\end{align}
with $T_{time}$ as the time-ordering operator. The effective Hamiltonian $\mathcal{H}_{eff}$ describes the dynamics of the system on the time scale much longer than a period $T$, thus the response is described well by an average over a period of oscillation. The matrix elements of the time-dependent Floquet Hamiltonian is \cite{zhou,zhouprb15,38,39,40,41}
\begin{eqnarray}
\mathcal{H}^{m,m^{'}}_{F}=\mathcal{H}_{0}\delta_{m,m^{'}}+m\hbar \omega \delta_{m,m^{'}}+\mathcal{H}^{'}_{m,m^{'}}
\end{eqnarray}
where $\mathcal{H}^{'}_{m,m^{'}}=V_{n}=\frac{1}{T}\int^{T}_{0}\mathcal{V}(t)e^{i(m-m^{'})t}dt=\frac{1}{T}\int^{T}_{0}\mathcal{V}(t)e^{in\omega t}dt,$ where $\mathcal{V}(t)$ is the time dependent periodic perturbation term.
Considering terms upto order  $1/\omega,$ the static time independent effective Hamiltonian is as follows
\begin{eqnarray}
\mathcal{H}_{eff}=\mathcal{H}_{0}+\frac{[V_{-1},V_{+1}]}{\hbar \omega}.
\label{eff-hamiltonian}
\end{eqnarray}

Note that all higher multi-photon state contributions vanish identically for our system.

\section{Modified Kubo Formula in the context of Floquet Theory}

Given a time dependent Hamiltonian $\mathcal{H}(t)$ influenced by a driving periodic potential $A_d(t) = A_d(T+t)$, with $T$ being the period, one can exploit the periodicity of the gauge potential to write the eigenstates of $\mathcal{H}(t)$ in the form \cite{PHG}

\beq
\ket{\Psi (t)} = e^{-i \epsilon_\alpha t} \ket{\Phi_\alpha (t)}.
\eeq

The states $\ket{\Phi (t)}$ are called Floquet states (analogous to the Bloch states for spatially periodic potentials) and they satisfy $\ket{\Phi (t+T)} = \ket{\Phi (t)}$. These states also satisfy the Schrodinger equation with $\epsilon_\alpha$'s being the quasi-energies.

\begin{align} \label{111}
\mathcal{H}(t) \ket{\Phi (t)} &= \epsilon_\alpha \ket{\Phi (t)}.
\end{align}

The Floquet states are orthonormal under a time averaged inner product defined as

\beq
\langle \braket{\Phi_\alpha (t) | \Phi_\beta (t)} \rangle := \frac{1}{T} \int_0^T dt  \braket{\Phi_\alpha (t) | \Phi_\beta (t)} = \delta_{\alpha \beta}.
\eeq

When applied to a system driven by an off-resonant optical field (i.e. the frequency $\omega$ of the field is larger than the bandwidth), real processes of photon absorption and emission cannot occur - this is why we stick to the high frequency regime, so that there is no dissipation. However, the off-resonant light can affect the system via virtual photon processes as described in  \cite{FWSM}. The physics of such virtual processes can be captured order by order by Fourier transforming the Floquet states to momentum space. 

\beq \label{2000}
\ket{\Phi (t)} = \sum_m e^{-im \Omega t} \ket{u_\alpha^m}
\eeq

In the equation above, $m$ is the Floquet index and it describes the order of the virtual photon process, while $\alpha$ is the band index for the full Hamiltonian. The states $\ket{u_\alpha^n}$ are related to the Floquet Hamiltonian as \cite{PHG}

\beq 
\sum_n \mathcal{H}_F^{mn} \ket{u_\alpha^n} = (\epsilon_\alpha  +m\Omega ) \ket{u_\alpha^m},
\eeq

with the Floquet Hamiltonian matrix elements defined as
\beq
\mathcal{H}_F^{mn} = \frac{1}{T} \int_0^T dt \mathcal{H} (t) e^{-i(m-n) \Omega t}.
\eeq

The full Floquet Hamiltonian has a Sambe space \cite{FWSM} (equivalent to a Hilbert space but used in context of Floquet theory) given by ${\bf F = \Omega \otimes H_{2x2}}$, where ${\bf H_{2x2}}$ is the Hilbert space of the undriven Hamiltonian, and ${\bf \Omega}$ is the space corresponding which captures the different virtual photon processes. This Hamiltonian can be approximated to linear order in perturbation theory in the high frequency expansion (HFE) as $\mathcal{H}^F_{eff} = H_{0,0} + \frac{1}{\Omega} [\mathcal{H}^{-10},\mathcal{H}^{10}]$, as used in our manuscript. The point of the effective Floquet Hamiltonian is that it shares the same quasi-energies and states of the full Floquet Hamiltonian to leading order in perturbation theory. The eigenstates of the effective Floquet Hamiltonian are defined as 
\beq
\mathcal{H}_{eff}^F \ket{e_\alpha} = \epsilon_\alpha \ket{e_\alpha},
\eeq 

and they are related to the Floquet states $\ket{u_\alpha^n}$ as 

\beq \label{1000}
\ket{e_\alpha} = \sum_n \ket{u_\alpha^n},
\eeq

as shown in \cite{awe}. The authors of \cite{PHG} state the form of the time-averaged conductivity tensor $\sigma_{ab}$ in the form of the Kubo formula, modified for the Floquet states in Eqn. (\ref{111}) as

\begin{align} \label{333}
\sigma_{ab} &= i \int \frac{d{\bf k}}{(2\pi)^d} \sum_{\alpha \neq \beta} \frac{f_\beta({\bf k}) - f_\alpha ({\bf k})}{\epsilon_\beta ({\bf k}) - \epsilon_\alpha ({\bf k})} \times \frac{\langle \braket{\Phi_\alpha  ({\bf k}) | J_b | \Phi_\beta  ({\bf k}) } \rangle \langle \braket{\Phi_\beta  ({\bf k}) | J_a | \Phi_\alpha  ({\bf k}) } \rangle}{\epsilon_\beta ({\bf k}) - \epsilon_\alpha ({\bf k}) + i\eta }
\end{align}

This is very similar to the Kubo formula with the following modifications: The energies have been replaced with the Floquet quasi-energies and the current correlation functions are time averaged. The $f$'s are the Fermi distribution function which take on a non-universal character out of equilibrium. Given the system described in our work, the current operator defined as $J =\frac{\partial \mathcal{H} (t)}{\partial A^\mu} = \hbar C_s \delta_{\mu z} + s \hbar v \sigma_\mu$ is independent of time. We evaluate the current correlation function above using the Fourier decomposition of Eqn.(\ref{2000}) as 

\begin{align} \label{222}
\langle \braket{\Phi_\beta  ({\bf k}) | J | \Phi_\alpha  ({\bf k}) } &= \frac{1}{T} \int_0^T dt \sum_m \sum_n e^{-i \Omega (n-m)t} \braket{u_\alpha^m|J|u_\beta^n} \nonumber \\ &= \sum_m \sum_ n \delta_{nm} \braket{u_\alpha^m|J|u_\beta^n} \nonumber \\ & = \sum_n \braket{u_\alpha^n|J|u_\beta^n},
\end{align}

where the simplification arises because the time dependent parts factor out.  \\

As derived in \cite{awe} (section III), in the HFE we find that $\ket{u_\alpha^n} \sim \mathcal{O} (\omega^{-n})$, and so to leading order only the zeroth level Floquet states $\ket{u^0_\alpha}$ contribute. Now, $\ket{u_\alpha^0}$ is given in terms of the eigenstates of the effective Floquet Hamiltonian from Eqn.(\ref{1000}), and thus the current correlator in Eqn. (\ref{222}) can be re-expressed as
\beq
\langle \braket{\Phi_\beta  ({\bf k}) | J | \Phi_\alpha  ({\bf k}) } = \sum_n \braket{u_\alpha^n|J|u_\beta^n} = \braket{u_\alpha^0|J|u_\beta^0} = \braket{e_\alpha|J|e_\beta}.
\eeq

The main point here is that the expectation value of observables are correctly computed using the eigenstates of the effective Hamiltonian to leading order in perturbation theory. This brings the expression for the conductivity tensor on the r.h.s. of Eqn.(\ref{333}) to exactly the Kubo form for the undriven case with the use of effective Floquet states and quasi-energies.

\begin{align} \label{444}
\sigma_{ab} &= i \int \frac{d{\bf k}}{(2\pi)^d} \sum_{\alpha \neq \beta} \frac{f_\beta({\bf k}) - f_\alpha ({\bf k})}{\epsilon_\beta ({\bf k}) - \epsilon_\alpha ({\bf k})} \times \frac{ \braket{e_\alpha  ({\bf k}) | J_b | e_\beta  ({\bf k}) } \braket{e_\beta  ({\bf k}) | J_a | e_\alpha  ({\bf k}) }}{\epsilon_\beta ({\bf k}) - \epsilon_\alpha ({\bf k}) + i\eta }
\end{align}

Since the Kubo formula as stated above is identically expressed using the Matsubara Green's function approach, using the effective Floquet states and quasi-energies in the Matusbara formalism yeilds the same conductivities as the one computed using Eqn.(\ref{444}). In the undriven case, the Matsubara formalism uses the Green's function derived from the Hamiltonian, and here it will correspondingly require the use of the Green's function of the effective Floquet Hamiltonian. \\

We note that the Fermi distribution is not universal for systems which are out of equilibrium but for some cases of system-bath couplings \cite{Eqb}, the steady state distribution becomes thermal, and we restrict our results to such cases. Additionally, the electrode chemical potential will be tiny in the regime of linear response as compared to the intrinsic chemical potential of the system, and so we ignore the electrode chemical potential in our calculations. This allows us to write the chemical potential in the Kubo formalism as a constant i.e. without accounting for sources at the boundary. \\

Note also that one could have conducted the calculation above using an additional slowly varying gauge field, which is then subsequently set to zero as we take the zero-frequency limit in the Kubo formula (i.e. the linear response regime), as done by the authors of \cite{PHG}

\section{Hall Conductivity Computation using modified Kubo Formalism}

The modified form of the Kubo formula as applicable to Floquet states of a strong and periodically driven system (derived in appendix B) is used in this appendix to compute the analytical form of the zero-temperature time-averaged components of the conductivity tensor.  The time-averaged anomalous Hall conductivity for the tilted WSM under the action of the circularly polarized light may now be derived from the zero frequency and zero wave-vector limit (i.e. the limit of an infinitesimal d.c. bias) of the current-current correlation function, constructed using the Matsubara Green's function method (with $\hbar=1$):
\begin{eqnarray}
\Pi_{ij}(\Omega,\mathbf{q}) &=& T\sum_{\omega_n}\sum_{s=\pm}\int \frac{d^3k}{(2\pi)^3} J_{i}^{(s)} G_{s}(i\omega_n,\mathbf{k})J_{j}^{(s)} G_{s}(i\omega_n-i\Omega_m,\mathbf{k}-\mathbf{q})\bigg|_{i\Omega_m\rightarrow \Omega + i\delta},
\end{eqnarray}
where $i,j=\{x,y,z\}$, $T$ is the temperature (setting the Boltzmann constant as unity)  and $\omega_n$($\Omega_m$) are the fermionic(bosonic) Matsubara frequencies. Here $G_{s}(i\omega_n,\mathbf{k})$ is the single particle Green's function of the electron and $J_{i}^{(s)} = e  \left(C_{s}\delta_{iz} +sv \sigma_i \right)$ is the current operator with $\delta_{ij}$ as the Kronecker delta.  One can relate the Hall conductivity to  the current-current correlation function as follows,
\begin{eqnarray}
\sigma_{xy} = -\lim_{\Omega\rightarrow 0}\frac{\Pi_{xy}(\Omega,0)}{i\Omega}.
\end{eqnarray}
The one-particle Green functions have the following form
\begin{equation}
G_{s}(i\omega_n,\mathbf{k}) = \frac{1}{2}\sum_{t=\pm1} \frac{1-st\bm\sigma \cdot \frac{\mathbf{k}-s(Q+\Delta)\mathbf{e}_z}{|\mathbf{k}-s(Q+\Delta)\mathbf{e}_z|}}{i\omega_n+\mu -C_{s}(k_z-s(Q+\Delta) +t v |\mathbf{k}-s(Q+\Delta)\mathbf{e}_z| -sC_s \Delta},
\end{equation}

where $\mu$ is the chemical potential. We sum over the Matsubara fermion frequencies and trace over Pauli $\sigma$-matrices to obtain the following form
\begin{equation}
\Pi_{xy}(\Omega,0) = \Pi_{xy}^{(+)}(\Omega,0)+\Pi_{xy}^{(-)}(\Omega,0),
\end{equation}
where we have separated the contributions from the two Weyl cones
\begin{gather} 
\Pi_{xy}^{(s)}(\Omega,0)= \Pi_0^{(s)}(\Omega,0)+ \Pi_{\mathrm{FS}}^{(s)}(\Omega,0),\\
\nonumber
\Pi_{0}^{(s)}(\Omega,0)= -se^2 \int_{-\Lambda_0 -s(Q+\Delta)}^{\Lambda_0 - s(Q+\Delta)} \frac{dk_z}{2\pi}\int_0^{\infty} \frac{k_{\perp}dk_{\perp}}{2\pi} \frac{2v^2 \Omega_m}{\Omega_m^2+4v^2k^2} \\ \label{9}
\times
\frac{k_z}{k}\bigg|_{i\Omega_m\rightarrow \Omega + i\delta},\\
\nonumber
\Pi_{\mathrm{FS}}^{(s)}(\Omega,0)= se^2 \int_{-\Lambda -s(Q+\Delta)}^{\Lambda - s(Q+\Delta)} \frac{dk_z}{2\pi}\int_0^{\infty} \frac{k_{\perp}dk_{\perp}}{2\pi} \frac{2v^2 \Omega_m}{\Omega_m^2+4v^2k^2}\\
\times
\frac{k_z}{k} \bigg \{ n_{F}(C_{s}k_z+vk-\mu+sC_{s}\Delta)-  n_{F}(C_{s}k_z -vk-\mu+sC_{s}\Delta)+1
 \bigg\}\bigg|_{i\Omega_m\rightarrow \Omega + i\delta}.
\end{gather}
$\Pi_{0}$ denotes the vacuum contribution for $\mu=0$, whereas $\Pi_{\mathrm{FS}}$ is the contribution of the states at the Fermi surface. 
$n_{F}(E) = (e^{(E-\mu)/T}+1)^{-1}$ is the Fermi distribution function and $k = \sqrt{k_z^2+k_{\perp}^2}$. The cut-off $\Lambda_0,$ which is introduced in the $k_z$ integral, is known not to affect the vacuum contribution to the Hall conductivity. However, the other cutoff in $\Pi_{FS}^{s}, $which is denoted as $\Lambda$, is crucial for finite Fermi surface effects in both the type-I and type-II regime.

Using eqn. (\ref{9}), we have
\begin{gather}
\sigma_{xy}^{(s)}= \sigma_0^{(s)}+ \sigma_{\mathrm{FS}}^{(s)},\\
\sigma_0^{(s)}= -e^2 \int_{-\Lambda_0 -s(Q+\Delta)}^{\Lambda_0 - s(Q+\Delta)} \frac{dk_z}{2\pi}\int_0^{\infty} \frac{k_{\perp}dk_{\perp}}{2\pi}\frac{sk_z}{2k^3},\\
\nonumber
\sigma_{\mathrm{FS}}^{(s)}= e^2 \int_{-\Lambda -s(Q+\Delta)}^{\Lambda - s(Q+\Delta)} \frac{dk_z}{2\pi}\int_0^{\infty} \frac{k_{\perp}dk_{\perp}}{2\pi}\\
\times
\frac{sk_z}{2k^3} \bigg[n_{F}(C_{s}k_z+vk-\mu+sC_{s}\Delta)-  n_{F}(C_{s}k_z -vk-\mu+sC_{s}\Delta)+1\bigg].
\end{gather}
Importantly, one should note that $sk_z/2k^3$ is the $z$-component of the Berry curvature of the Weyl cone with chirality $s$. Interestingly, both the tilt $C_s$ and the Floquet parameter $\Delta$ term has no effect on the Berry curvature component, but only affect the Fermi-Dirac distribution function.

Taking the $T\rightarrow 0$, and performing the $k_{\perp}$ integration, we get
\begin{gather}
\sigma_0^{(s)}= -\frac{se^2}{8\pi^2}\int_{-\Lambda_0 -s (Q+\Delta)}^{\Lambda_0 -s (Q+\Delta)} dk_z\,\mathrm{sign}(k_z),\\
\nonumber
\sigma_{\mathrm{FS}}^{(s)}=-\frac{se^2}{8\pi^2}\int_{-\Lambda -s (Q+\Delta)}^{\Lambda -s (Q+\Delta)} dk_z\left[\mathrm{sign}(k_z)-\frac{vk_z}{|C_{s}k_z-\mu+sC_{s} \Delta |}\right]\\
\times\left[(\Theta\left(v^2 k_z^2 - (C_{s}k_z +sC_{s}\Delta-\mu)^2\right)-1\right],
\label{eq:sigmaxy12kz}
\end{gather}
where $\Theta(x)$ is the Heaviside function. 
\begin{gather}
\sigma_{xy}=\sigma_0+ \sigma_{\mathrm{FS}}\\
\sigma_0= -\frac{e^2}{4\pi^2}\int_{-\Lambda_0 - (Q+\Delta)}^{\Lambda_0 - (Q+\Delta)} dk_z\,\mathrm{sign}(k_z) = \frac{e^{2}}{2\pi^{2}}(Q+\Delta)
\end{gather}

Noticeably, the optical field has lead to a positive offset to $\sigma_{0}$. In this case the Fermi surface contribution to the Hall effect can be written as
\begin{align}\label{22}
\sigma_{FS}^{s}=\frac{-se^{2}}{8\pi^{2}}\int_{-\Lambda -s (Q+\Delta)}^{\Lambda -s (Q+\Delta)} dk_z\left[\mathrm{sign}(k_z)-\frac{vk_z}{|C_{s}k_z-\mu+sC_{s} \Delta |}\Big[\theta \Big(v^{2}k_{z}^{2}-\Big(C_{s}k_{z}+sC_{s}\Delta-\mu\Big)^{2}\Big)-1\Big]\right]
\end{align}

The integral can be evaluated for both types of WSMs as stated in the main text.

\section{Schematic Design for Experimental Realization}

\begin{figure} [h]
\fbox{\includegraphics[scale=.8]{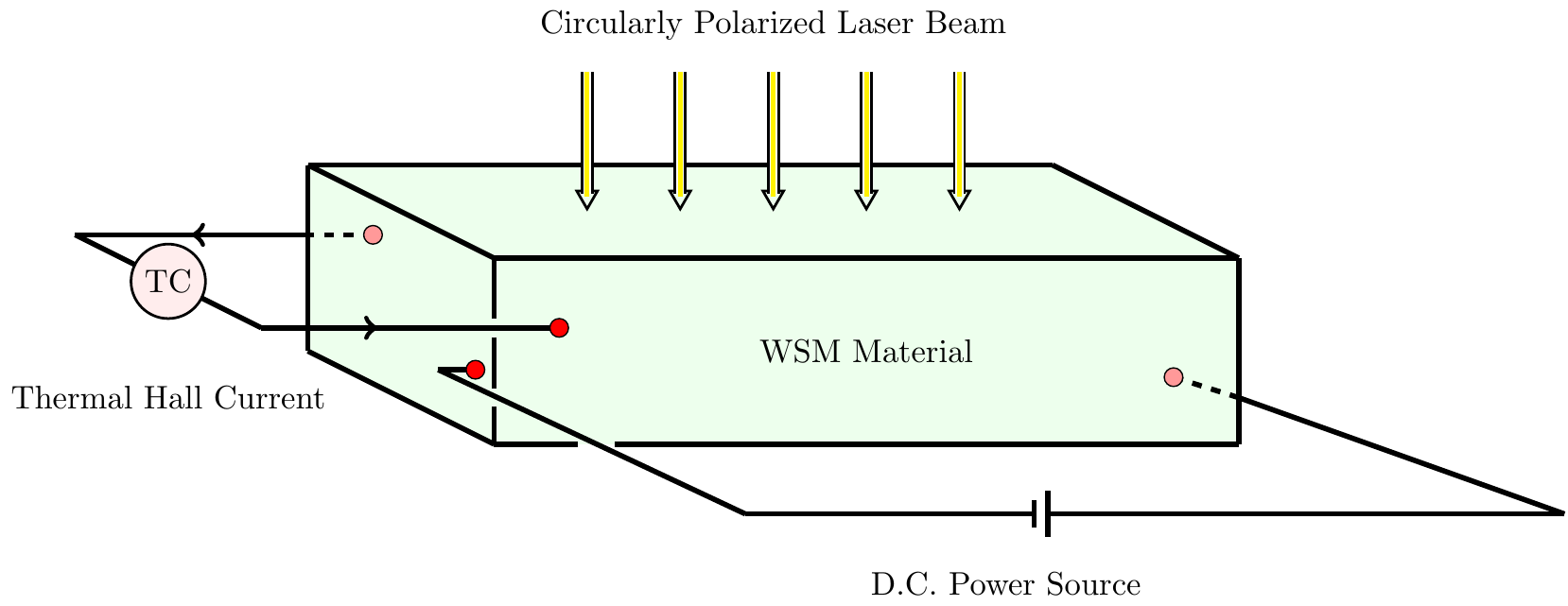}}
\caption{A schematic for an pump-probe experiment to measure the anomalous thermal Hall conductivity of a WSM sample. TC represents a thermocouple which may be used to determine the temperature gradient which maps to the thermal Hall current.}
\end{figure}

We propose the use of the experimental setup in Fig. 1 (of the appendix) above to test the validity of our results, where the WSM sample is connected to two metallic leads which introduce a small d.c. bias (consistent with the use of linear response theory), and the thermocouple measures the transverse temperature gradient which is directly related to the thermal Hall current \cite{Zyu}. The circularly polarized irradiation field should possess a frequency governed by $-\omega/2 < E < \omega/2$, where E represents the band energy close to the Weyl points, i.e. in the linear dispersion regime. The purpose of the d.c. source is to induce a transverse Hall, and consequentially, an anomalous thermal Hall current.  \\ \\
In principle, one can design a pump-probe experiment, which serves the dual purpose of driving the system into a non-equilibrium state with high temporal resolution, as well as measuring the transport properties of the WSM sample. Note that the frequency of the pump pulse ($\omega_{\textrm{pump}}$) is usually significantly smaller than the frequency of the optical field ($\omega$), i.e. $\omega_{\textrm{pump}} << \omega$, and the high frequency $\omega$ implies that the Floquet bands are well separated \cite{Sci}, additionally increasing the precision of the high frequency expansion. For a choice of polarization of the optical field, the dispersion relation for this system can be verified using {\it time-resolved} angle-resolved photoemission spectroscopy (tr-ARPES \cite{HRK,tr}. It is also possible to directly measure the Nernst and anomalous thermal Hall conductivities by conventional transport experiments \cite{A,B,C}, using the two distinct classes of WSMs used as samples.